\documentclass[prl,twocolumn,amsmath,amssymb]{revtex4}

\usepackage{ulem}
\usepackage{graphicx}
\usepackage{dcolumn}
\usepackage{bm}

\usepackage{braket}
\usepackage{cases}

\usepackage{xcolor}

\begin{document}

\title{Magnetic phase transition in a mixture of two interacting superfluid Bose gases at finite temperature}

\author{Miki Ota, Stefano Giorgini, and Sandro Stringari}

\affiliation{INO-CNR BEC Center and Dipartimento di Fisica, Universit\`a di Trento, 38123 Trento, Italy}

\date{\today}

\begin{abstract}
The miscibility condition for a binary mixture of two interacting Bose-Einstein condensates is shown to be deeply affected by interaction driven thermal fluctuations. These give rise to a first order phase transition to a demixed phase with full spatial separation of the two condensates, even if the mixture is miscible at zero temperature. Explicit predictions for the isothermal compressibility, the spin susceptibility, and the phase transition temperature $T_M$ are obtained in the framework of Popov theory, which properly includes beyond mean-field quantum and thermal fluctuations in both the spin and density channels. For a mixture of two sodium condensates occupying the hyperfine states $\ket{F=1,m_F=1}$ and $\ket{F=1,m_F=-1}$ respectively, $T_M$ is predicted to occur at about $0.7$ times the usual BEC critical temperature.
\end{abstract}

\maketitle



\textit{Introduction.---}The miscibility of liquids and gases, and in particular its temperature dependence, is a topic of high relevance in the study of classical fluids \cite{Petrucci}. For quantum mixtures, this question was addressed long time ago in the context of ${}^3\mathrm{He}$-${}^4\mathrm{He}$ liquids \cite{Baym1978}, and more recently for mixtures of quantum gases \cite{book, Pethick2001}. In particular, weakly interacting binary Bose  gases occupying two different hyperfine states are the simplest, yet interesting example of quantum mixtures, for which the problem of miscibility has been intensively investigated, both experimentally \cite{Hall1998, Papp2008, Mccarron2011, Tojo2010, Nicklas2011} and theoretically \cite{Ho1996, Pu1998, Ao1998, Trippenbach2000, Ohberg1998, Chien2012, Roy2015, Armaitis2015, Lee2016, Boudjemaa2018, Shi2000, Schaeybroeck2013}. The  theoretical studies have revealed that, in the zero temperature mean field regime, the mixture is stable against phase separation if the inequality $g_{12}^2 < g_{11} g_{22}$ holds, where $g_{ij}$ is the coupling constant for the intra-species ($g_{11}$ and $g_{22}$) and inter-species ($g_{12}$) interactions \cite{book, Ho1996, Pu1998, Ao1998, Trippenbach2000, Chien2012}. At finite temperature, theoretical studies have mainly focused on harmonically trapped systems, by means of the Hartree-Fock \cite{Shi2000, Ohberg1998, Schaeybroeck2013}, Zaremba-Nikuni-Griffin \cite{Lee2016} and Hartree-Fock-Bogoliubov \cite{Armaitis2015, Roy2015, Boudjemaa2018} theories. Although they differ in the treatment of the intra-species interaction, all the above approaches treat the inter-species coupling at the mean-field level, thereby providing an inaccurate description of the thermal fluctuations associated with the spin degree of freedom.
\par
In this Letter we study the case of a uniform bosonic mixture occupying two different hyperfine states, satisfying  the miscibility condition at zero temperature. Contrary to intuitive arguments based on entropy considerations, we predict the occurrence of a peculiar magnetic phase transition at finite temperature, characterized by the remarkable space separation of the condensate components of the two Bose gases. Our theoretical approach is based on the generalization of Popov theory \cite{Popov1987, Fedichev1998, Giorgini2000} applied to a mixture of two interacting Bose gases. This theory  properly accounts for quantum and thermal fluctuations in both the density and spin channels, beyond the mean-field approach. Differently from the Hartree-Fock approximation, which is shown to predict a magnetic phase transition as a consequence of the negative value of the spin susceptibility at a finite temperature below the Bose-Einstein condensation (BEC) critical point, the phase transition predicted by Popov theory is not related to the occurrence of a dynamic instability and has a first-order nature arising from an energetic instability. If the inter-species interaction is close to, but still smaller than the intra-species value ($0<\delta g=\sqrt{g_{11}g_{22}}-g_{12} \ll \sqrt{g_{11}g_{22}}$), the predicted magnetic transition is found to occur significantly below the BEC critical temperature.



\par
\textit{Compressibility and susceptibility.---} We start from the investigation of the isothermal compressibility $\kappa_T$ and magnetic susceptibility $\kappa_M$, related, respectively, to the density and spin response of the system \cite{Stringari2009}. For mixtures of two gases, the mixed state is dynamically stable against density and spin fluctuations if $\kappa_T$ and $\kappa_M$ are positive \cite{Landau1980, Viverit2000}. The two quantities are defined through the Helmoltz free energy $F$ per unit volume $V$ according to:
\begin{equation}\label{Eq.kappa}
\kappa_T = \left( \frac{\partial^2 F/V}{\partial n^2} \right)^{-1} \, , \quad
\kappa_M = \left( \frac{\partial^2 F/V}{\partial m^2} \right)^{-1}_{m=0} \, ,
\end{equation}
with $n=n_1+n_2$ the total density of particles and $m=n_1-n_2$ the magnetization density. The simplest and widely used theory to investigate the thermodynamic behavior of binary Bose gases is the Hartree-Fock (HF) model \cite{Shi2000, Ohberg1998, Schaeybroeck2013}. For mixtures where both components are in the condensate phase, the HF free energy takes the form:
\begin{align} \label{Eq.FHF}
\frac{F^\mathrm{HF}}{V} =& \frac{g}{2} \left( n_1^2 + n_2 ^2 \right) +g_{12} n_1 n_2 \nonumber \\
&+ g\frac{\zeta(3/2)^2}{\lambda_T^6} + \frac{1}{\beta V} \sum_i  \sum_\mathbf{k} \ln \left( 1 - e^{-\beta (\varepsilon_\mathbf{k} + g n_{i, 0})} \right) ,
\end{align}
with $\zeta (s)$ and $\lambda_T$ the Riemann zeta function and thermal de Broglie wavelength, respectively, and we have assumed $g_{11} = g_{22} = g$. In Eq. \eqref{Eq.FHF}, the first line corresponds to the $T=0$ mean-field contribution, while the second line accounts for the thermal contribution of the single-particle excitations, where $\varepsilon_\mathbf{k} = \hbar^2 k^2 / (2M)$ is the particle kinetic energy and $n_{i, 0} = n_i - \zeta(3/2)/\lambda_T^3$ is the Bose condensed particles density holding to the lowest order in the interaction. The HF model predicts BEC to occur when $n_{i, 0} = 0$, corresponding to the critical temperature $k_B T_{i,\mathrm{BEC}}=2\pi \hbar^2/M(n_i/\zeta(3/2))^{2/3}$. At zero temperature, one verifies that Eqs. \eqref{Eq.kappa} and \eqref{Eq.FHF} yields $\kappa_{T,M} = 2/(g \pm g_{12})$, so that phase separation occurs for $g \leq g_{12}$. For example, in the case of a mixture of ${}^{23}\mathrm{Na}$ atoms occupying the  hyperfine states $\ket{F=1,m_F=\pm 1}$, one has $\delta g/g=0.07$ yielding an increase of a factor $\sim 14$ of the $T=0$ value of the spin polarizability with respect to the value obtained in the absence of inter-species interactions. The huge increase of the spin susceptibility has been recently demonstrated  experimentally in the case of a harmonically trapped mixture of sodium atoms \cite{Bienaime2016, Fava2018}. 
\par
\begin{figure}[t]
\begin{center}
\includegraphics[width=0.8\columnwidth]{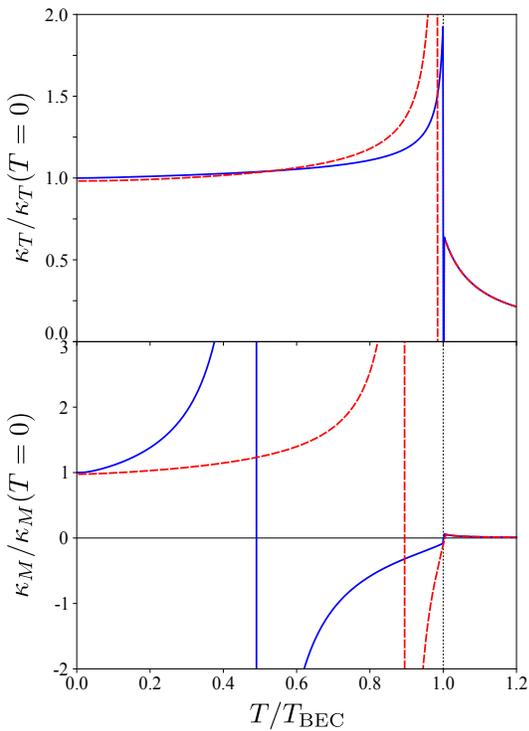}
\caption{Isothermal compressibility (a) and spin susceptibility (b) Eq. \eqref{Eq.kappa} for binary mixtures of Bose gases, with interaction parameters $gn/(k_B T_\mathrm{BEC})=0.1$ and $\delta g/g=0.07$. The blue solid and the red dashed lines are the predictions of HF theory (Eq. \eqref{Eq.FHF}) and Popov theory (Eq. \eqref{Eq.FPo}), respectively. Both quantities are normalized to the mean-field $T=0$ values, $\kappa_{T, M}(T=0) = 2/(g \pm g_{12})$.} 
\label{fig:thermo}
\end{center}
\end{figure}
In Fig. \ref{fig:thermo} we report the temperature dependence of the isothermal compressibility and of the spin susceptibility in the case of the sodium mixture discussed above, in the unpolarized configuration $n_1=n_2=n/2$. Remarkably, the susceptibility predicted by the HF theory shown in panel (b) exhibits a divergent behavior at $T \simeq 0.5 T_\mathrm{BEC}$, therefore signaling the onset of a magnetic dynamic instability. The origin of this instability can be understood if one writes the analytical expression for the spin susceptibility, obtained by expanding the Bose distribution function in the high-temperature regime $k_B T \gg gn$:
\begin{equation}\label{Eq.kappaM_HF}
2 \left( \kappa_M^\mathrm{HF} \right)^{-1} \simeq \delta g - g^{3/2} \frac{\sqrt{\pi}}{\lambda_T^3} \sqrt{\frac{\beta}{n_0}} \, . 
\end{equation}
The onset of the dynamical instability in the HF description is due to the last $g^{3/2}$-term in Eq. \eqref{Eq.kappaM_HF}, arising from interaction driven thermal fluctuations. As the temperature increases, beyond mean-field effects are enhanced, eventually leading to a divergent behavior of $\kappa_M^\mathrm{HF}$ at finite temperature. These results for the dynamical instability are consistent with the earlier work of Ref. \cite{Schaeybroeck2013}, in which the author calculates the grand-canonical potential in the HF approximation, and predicts the occurrence of a phase-separation of BEC components caused by thermal effects, even when $\delta g > 0$.
\par
As explicitly shown in Eq. \eqref{Eq.FHF}, the HF theory accounts for the inter-species interaction only to lowest order, linear in $g_{12}$. Since the divergence of the HF spin susceptibility arises from terms beyond linear order in the intra-species interaction $g$, it is natural to ask how this instability is modified by the inclusion of higher order terms in $g_{12}$. In order to answer this question, we develop the Popov theory for a mixture of two condensates, starting from a model Hamiltonian which treats in a consistent way both inter and intra-species interactions. The associated grand-canonical Hamiltonian $K$ can be diagonalized by means of Bogoliubov transformations, as well as proper renormalizations of the coupling constants \cite{Griffin1996}. The details of the derivation will be given in a subsequent paper, and here we only provide the final result:
\begin{equation}\label{Eq.K_Po}
K = \Omega_0 + \sum_{\mathrm{k} \neq 0} \left( E^+_{\mathbf{k}} \alpha_\mathbf{k}^\dagger \alpha_\mathbf{k} + E^-_{ \mathbf{k}} \beta_\mathbf{k}^\dagger \beta_\mathbf{k} \right) \, ,
\end{equation}
where $\alpha_\mathbf{k}^\dagger$ and $\beta_\mathbf{k}^\dagger$ are, respectively, the creation operators for the quasiparticles in the density and spin channels obeying Bose statistics, and $\Omega_0$ is the thermodynamic potential of the vacuum of these quasiparticles. The excitation spectrum of the system reads $E^\pm_\mathbf{k} = \sqrt{\varepsilon_\mathbf{k}^2 + 2 \Lambda_\pm \varepsilon_\mathbf{k}}$ where,
\begin{equation}\label{Eq.Lambda}
\Lambda_\pm = \frac{1}{2} \left( g n_0 \pm \sqrt{(g^2-g_{12}^2) m_0^2 + g_{12}^2 n_0^2} \right) \, ,
\end{equation}
are the Bogoliubov sound velocities, with $n_0 = n_{1,0} + n_{2,0}$ and $m_0 = n_{1,0} - n_{2,0}$. Eventually one obtains the expression for the free energy using the thermodynamic relation $F = \Omega + \sum_i \mu_i n_i$, with $\Omega = \beta^{-1}\ln \left( \mathrm{Tr} e^{-\beta K} \right)$ and $\mu_i$ the chemical potential for the $i^\mathrm{th}$ component evaluated from the saddle point equation $\partial \Omega / \partial n_{i,0} = 0$. In the BEC mixed phase, where both components are condensed, one finds:
\begin{align} \label{Eq.FPo}
\frac{F}{V} =& \frac{g}{2} \left( n_1^2 + n_2 ^2 \right) +g_{12} n_1 n_2 \nonumber \\
&+ g \frac{\zeta(3/2)^2}{\lambda_T^6} + \frac{1}{\beta V} \sum_\pm  \sum_\mathbf{k} \ln \left( 1 - e^{-\beta E^\pm_\mathbf{k}} \right) \nonumber  \\
&+ \left( \frac{M}{2\pi\hbar^2} \right)^{3/2} \frac{4}{15 \sqrt{\pi}} \sum_\pm \left( 2 \Lambda_\pm \right)^{5/2} \, .
\end{align}
In this expression, the first two lines are similar to the ones of HF theory in Eq. \eqref{Eq.FHF}, the thermal contribution of single-particles being now replaced by that of quasiparticles. As for the last line of Eq. \eqref{Eq.FPo} it corresponds to the contribution arising from quantum fluctuations. At $T=0$ this term reduces to the Lee-Huang-Yang energy functional, that was used in Ref. \cite{Petrov2015} to predict the existence of self-bound quantum droplets, as a result of the competition between mean-field attraction and beyond mean-field repulsion \cite{Cabrera2018, Semeghini2018}. 
\par 
Results for the response functions Eq. \eqref{Eq.kappa} are shown as red dashed lines in Fig. \ref{fig:thermo}. In panel (a), the isothermal compressibility is found to lie close to the HF prediction in a wide range of temperatures below $T_\mathrm{BEC}$.  However as shown in panel (b), we find that the spin susceptibility predicted by the Popov theory deviates strongly from the HF calculation. For $T>T_\mathrm{BEC}$, Popov theory reduces to the HF theory. In order to understand the major differences provided by the two approaches, we derive the high-temperature analytical expression of the spin susceptibility, now calculated within the Popov theory Eq. \eqref{Eq.FPo}. We find:
\begin{align}
2 \left( \kappa_M\right)^{-1} \simeq & \delta g - g^{3/2} \frac{\delta g}{g_{12}} \frac{2 \sqrt{\pi}}{\lambda_T^3} \sqrt{\frac{\beta}{n_0}}  \nonumber \\
& \times \left[ \left( 1 + \frac{g_{12}}{g} \right)^{3/2} - \left( 1 + \frac{g_{12}}{g} \right) \sqrt{\frac{\delta g}{g}} \right] \, . \label{Eq.kappaM_Po}
\end{align}
In contrast to the HF prediction Eq. \eqref{Eq.kappaM_HF}, the Popov approach gives rise to terms proportional to $\delta g$ also for the beyond mean-field terms (second term in the right-hand side of Eq. \eqref{Eq.kappaM_Po}). A careful analysis of the grand-canonical Hamiltonian in Eq. \eqref{Eq.K_Po} reveals that the emergence of such beyond mean-field terms in $g_{12}$ is due to the correct treatment of the two-component anomalous densities $\langle \psi_i^\dagger \psi_j \rangle_{i \neq j}$ and $\langle \psi_i \psi_j \rangle_{i \neq j}$, with $\psi_i$ the bosonic field operator for the particles in the component $i$. These anomalous averages are natural extensions of the single component anomalous density $\langle \psi_i \psi_i \rangle$ of Bogoliubov theory \cite{Griffin1996}, and are the consequences of presence of Bose-Eistein condensation in both components. The inclusion of such terms in the grand-canonical Hamiltonian Eq. \eqref{Eq.K_Po} is crucial to provide a proper description of both spin and density fluctuations. It is worth noticing that besides the well-known solution $\delta g = 0$, Eq. \eqref{Eq.kappaM_Po} possesses a second root, leading to a dynamical instability at finite temperature even if $\delta g > 0$. In Fig. \ref{fig:thermo}(b) the divergence of $\kappa_M$ is found to occur at $T \sim 0.9 T_\mathrm{BEC}$. However at this temperature the system is already phase-separated, as we discuss in the next section.
 


\par
\textit{Magnetic phase separation.---} In the previous section, we have established the region where the mixed binary configuration is dynamically stable. We now turn to the investigation of a possible energetic instability, associated with the emergence of an energetically favorable  phase separated state. Let us consider an unpolarized Bose mixture, miscible at zero temperature ($\delta g > 0$). Since we consider a uniform system, the mixture is prone to separate into two domains (A, B) of equal volume $V/2$, conserving the total density $n^A = n^B = n$, but with opposite magnetization $m^A = -m^B=m$. The two domains are in equilibrium when both the pressure ($P^A = P^B$) and the chemical potential ($\mu_i^A = \mu_i^B$) equilibrium conditions are satisfied. While the equilibrium  condition for the pressure is automatically satisfied for the symmetric configuration considered here, the chemical potential equilibrium is found to be fulfilled only if, in each domain one of the two components is in the normal phase. For such a configuration the Popov free energy in each domain is given by:
\begin{align}\label{Eq.FPo_Sep}
\frac{F}{V} =& \frac{g}{2} \left( n_1^2 + 2 n_2^2  + \frac{\zeta(3/2)^2}{\lambda_T^6} \right) +g_{12} n_1 n_2 + \mu_2^\mathrm{IBG} n_2 \nonumber \\
&+ \left( \frac{M}{2\pi\hbar^2} \right)^{3/2} \frac{4}{15 \sqrt{\pi}} \left( 2 g n_{1,0} \right)^{5/2} \nonumber \\
&+ \frac{1}{\beta V} \sum_\mathbf{k} \ln \left( 1 - e^{-\beta E_\mathbf{k}} \right) \nonumber \\ 
&+ \frac{1}{\beta V} \sum_\mathbf{k} \ln  \left( 1 - e^{-\beta (\varepsilon_\mathbf{k} - \mu_2^\mathrm{IBG} )} \right) \, ,
\end{align}
where we chose $n_2$ to be the minority component in the normal phase. The ideal Bose gas chemical potential $\mu_2^\mathrm{IBG}$ is defined through the relationship $n_2 = g_{3/2}(e^{\beta \mu_2^\mathrm{IBG}})/\lambda_T^3$, with $g_p(z)$ the usual Bose special function \cite{book}. As for the majority component in the condensed phase, it is now described by the quasi-particle energy $E_\mathbf{k} = \sqrt{\varepsilon_\mathbf{k}^2 + 2 \varepsilon_\mathbf{k} g n_{1,0}}$.
\begin{figure}[t]
\begin{center}
\includegraphics[width=0.8\columnwidth]{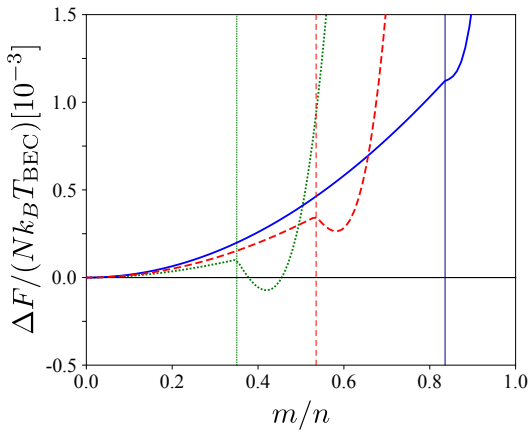}
\caption{Difference of free energies between the miscible state ($m=0$) and the phase-separated state described in the main text, for $gn/(k_B T_\mathrm{BEC})=0.1$ and $\delta g/g=0.07$. Blue solid line: $T<T^*$, red dashed line: $T^*<T<T_M$, green dotted line $T>T_M$. The vertical lines indicate the critical magnetization $m=n-2 \zeta(3/2)/\lambda_T^3$ above which the minority component is purely thermal.} 
\label{fig:F}
\end{center}
\end{figure}
\par
\begin{figure}[t]
\begin{center}
\includegraphics[width=0.8\columnwidth]{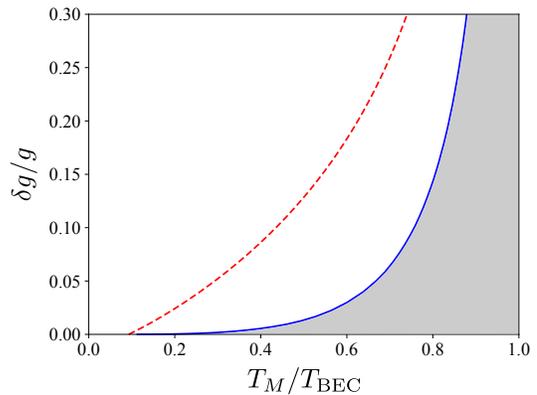}
\caption{Phase diagram for binary condensates with $gn/(k_B T_\mathrm{BEC}) = 0.1$. The blue solid and the red dashed lines are the phase transition temperature $T_M$, and characteristic temperature $T^*$, respectively. The gray area corresponds to the regime of phase-separation.} 
\label{fig:phaseDiag}
\end{center}
\end{figure}

Figure \ref{fig:F} shows the calculated free energy as a function of the magnetization density, for different values of temperature. At low temperature, the free energy is a monotonously increasing function (see blue solid line), with a unique minimum at zero magnetization, corresponding to the mixed state. At a given temperature hereafter called $T^*$, a second minimum starts to develop in the region where the minority component is purely thermal, $m > n-2\zeta(3/2)/\lambda_T^3$ (red dashed line). As already stressed, the emergence of such metastable state corresponds to the fulfillment of the chemical potential equilibrium between the two domains. An analytical expression for the temperature $T^*$ can be obtained from Eq. \eqref{Eq.FPo_Sep}, by employing the high temperature $k_B T \gg gn$ expansion for the Bose distribution function:
\begin{equation}\label{Eq.T*}
\frac{T^*}{T_\mathrm{BEC}} \simeq \frac{\delta g}{g} \frac{\zeta(3/2)}{\sqrt{2 \pi}} \sqrt{\frac{T_\mathrm{BEC}}{gn}} \, .
\end{equation}
By further increasing the temperature the energy of the metastable state decreases, eventually reaching the same energy as the unpolarized state, therefore signaling the onset of a first order phase transition. Hereafter we use the notation $T_M$ to denote this magnetic phase transition temperature, above which the mixed state is energetically unstable with respect to the phase separated state (green dotted line in Fig. \ref{fig:F}). The new equilibrium phase predicted by Popov theory is hence characterized by a full space separation of the Bose-Einstein condensed components of the two atomic species, their thermal components remaining instead mixed, with a finite magnetization. In Fig. \ref{fig:phaseDiag} we show the phase diagram of the two-component Bose mixture, by plotting the characteristic temperature $T^*$, providing the onset of a minimum in the free energy with $m\ne 0$ and the phase transition temperature $T_M$ as a function of $\delta g/g$. For the sodium mixture where $\delta g /g=0.07$, we find that the phase-separated state appears as a metastable state at $T^*=0.36 T_\mathrm{BEC}$ while  the phase transition occurs at $T_M = 0.71 T_\mathrm{BEC}$. We briefly note that as $\delta g / g \rightarrow 0$, $T^*$ tends to a finite value ($ \simeq 0.1 T_\mathrm{BEC}$), as a consequence of quantum fluctuations, in contrast with Eq. \eqref{Eq.T*} which only holds if $T^* \gg gn/k_B$. We also find that the phase separated state disappears slightly above the critical temperature $T_\mathrm{BEC}$. At this temperature, the mixture becomes again miscible with both components in the normal phase.
\par
Finally, let us briefly discuss the effects of inhomogeneity on the phase separation. So far, all the experiments on binary Bose gases at finite temperature have been performed in presence of harmonic potentials (see for example Ref. \cite{Fava2018}), and the phase transition discussed in this section has never been observed. This can be understood from the suppression of feedback between thermal and condensate atoms in a trapped gas. Indeed, in a trap the condensate atoms occupy the center of the cloud, while the thermal ones are spread out, and consequently the overlap between the two components is greatly reduced. From the aforementioned free energy analysis, we have verified that by neglecting the coupling between the condensate and the thermal atoms, the phase separation does not occur for any value of $T$ \cite{note1}. 



\par 
\textit{Conclusion.---} By developing a beyond mean-field theory, which properly includes thermal fluctuation effects in both the density and spin channels, we have investigated the thermodynamic behavior of a uniform binary mixture of weakly interacting Bose gases and predicted the occurrence of a first order magnetic transition. The phase transition is characterized by the space separation of the two condensates and the formation of polarized domains. Our analysis has revealed that the corresponding phase separation can occur, even if the  inequality $g_{12} < g$ is satisfied (and hence the mixture is miscible at zero temperature) as a consequence of interaction induced thermal fluctuations and of the coupling between the condensate and the thermal atoms. Recently, box-like potentials providing uniform trapping have become available for both Bose \cite{Lopes2017} and Fermi \cite{Mukherjee2017} gases. Thus, the experimental possibility of observing the predicted magnetic phase transition is a realistic option.  Important open issues concern the propagation of sound in these polarized domains, the possible emergence of a similar magnetic phase transition in two dimensions, the relevance of finite-size effects caused by the trapping and the structure of the interface between different domains.
\par
During the final preparation of the manuscript, we become aware of the  work of Ref. \cite{Hryhorchak}, where beyond mean-field effects are included by means of large-N expansion techniques. The resulting predictions for the behavior of spin susceptibility are consistent with our findings  based on Popov theory.
\acknowledgments
{\bf Acknowledgments}. This project has received funding from the EU Horizon 2020 research and innovation programme under grant agreement No. 641122 QUIC, and by Provincia Autonoma di Trento.





\begin{thebibliography}{99}
\bibitem{Petrucci} R. H. Petrucci, W.S. Harwood, G. E. Herring, \textit{General Chemistry Principles \& Modern Applications. 9th Edition} (Pearson, Toronto, 2007).
\bibitem{Baym1978} G. Baym and C. J. Pethick, in \textit{The Physics of Liquid and Solid Helium}, edited by K.H. Bennemann and J. Ketterson, (Wiley, New York, 1978), Part II, pp. 123-175.
\bibitem{book} L. Pitaevskii and S. Stringari, {\textit Bose-Einstein Condensation and Superfluidity} (Oxford University Press, Oxford, 2016).
\bibitem{Pethick2001} C. J. Pethick and H. Smith, {\it Bose-Einstein Condensation in Dilute Gases} (Cambridge University Press, Cambridge, 2002).
\bibitem{Hall1998} D.S. Hall, M.R. Matthews, J.R. Ensher, C.E. Wieman, and E.A. Cornell, Phys. Rev. Lett. \textbf{81}, 1539 (1998).
\bibitem{Papp2008} S. B. Papp, J.M. Pino, and C.E.  Wieman, Phys. Rev. Lett. \textbf{101}, 040402 (2008).
\bibitem{Mccarron2011} D.J. McCarron, H.W. Cho, D.L. Jenkin, M.P.  K{\"{o}}ppinger, and S.L. Cornish, Phys. Rev. A \textbf{84} 011603(R) (2011).
\bibitem{Tojo2010} S. Tojo, Y. Taguchi, Y. Masuyama, T. Hayashi, H. Saito, and T. Hirano, Phys. Rev. A \textbf{82} 033609 (2010).
\bibitem{Nicklas2011} E. Nicklas, H. Strobel, T. Zibold, C. Gross, B.A. Malomed, P.G. Kevrekidis, and M.K. Oberthaler, Phys. Rev. Lett. \textbf{107} 193001 (2011).
\bibitem{Ho1996} T.-L. Ho, and V.B. Shenoy, Phys. Rev. Lett. \textbf{77}, 3276 (1996).
\bibitem{Pu1998} H. Pu, and N.P. Bigelow, Phys. Rev. Lett. \textbf{80}, 1130 (1998).
\bibitem{Ao1998} P. Ao, and S.T. Chui, Phys. Rev. A \textbf{58}, 4836 (1998).
\bibitem{Trippenbach2000} M. Trippenbach, K. G{\'{o}}ral, K. Rzazewski, B. Malomed, and Y.S. Band, J. Phys. B \textbf{33}, 4017 (2000).
\bibitem{Chien2012} C.-C. Chien, F. Cooper, E. Timmermans, Phys. Rev. A \textbf{86}, 023634 (2012).
\bibitem{Ohberg1998} P. \"{O}hberg, and S. Stenholm, Phys. Rev. A \textbf{57}, 1272 (1998).
\bibitem{Shi2000} H. Shi, W.-M. Zheng, and S.-T. Chui, Phys. Rev. A \textbf{61}, 063613 (2000).
\bibitem{Schaeybroeck2013} V.B. Schaeybroeck, Physica A \textbf{392}, 3806 (2013).
\bibitem{Lee2016} K.L. Lee, N.B. J{\o}rgensen, I-K. Liu, L. Wacker, J.J. Arlt, and N.P. Proukakis, Phys. Rev. A \textbf{94}, 013602 (2016).
\bibitem{Roy2015} A. Roy, and D. Angom, Phys. Rev. A \textbf{92}, 011601(R) (2015).
\bibitem{Armaitis2015} J. Armaitis, H.T.C. Stoof, and R.A. Duine, Phys. Rev. A \textbf{91}, 043641 (2015).
\bibitem{Boudjemaa2018} A. Boudjem{\^{a}}a, Phys. Rev. A \textbf{97}, 033627 (2018).
\bibitem{Popov1987} V.N. Popov, \textit{Functional Integrals and Collective Excitations}, (Cambridge University Press, Cambridge, 1987).
\bibitem{Fedichev1998} P. O. Fedichev and G. V. Shlyapnikov, Phys. Rev. A \textbf{58}, 3146 (1998).
\bibitem{Giorgini2000} S. Giorgini, Phys. Rev. A \textbf{61}, 063615 (2000).
\bibitem{Stringari2009} S. Stringari, Phys. Rev. Lett. \textbf{102}, 110406 (2009).
\bibitem{Landau1980} L. D. Landau and E. M. Lifshitz, \textit{Statistical Physics, Part 1} (Pergamon, Oxford, 1980).
\bibitem{Viverit2000} L. Viverit, C. J. Pethick, and H. Smith, Phys. Rev. A \textbf{61}, 053605 (2000).
\bibitem{Bienaime2016} T. Bienaim{\'{e}}, E. Fava, G. Colzi, C. Mordini, S. Serafini, C. Qu, S. Stringari, G. Lamporesi, G. Ferrari, Phys. Rev. A \textbf{94}, 063652 (2016).
\bibitem{Fava2018} E. Fava, T. Bienaim{\'{e}}, C. Mordini, G. Colzi, C. Qu, S. Stringari, G. Lamporesi, and G. Ferrari, Phys. Rev. Lett. \textbf{120}, 170401 (2018).
\bibitem{Griffin1996} A. Griffin, Phys. Rev. B \textbf{53}, 9341 (1996).
\bibitem{Petrov2015} D.S. Petrov, Phys. Rev. Lett. \textbf{115}, 155302 (2015).
\bibitem{Cabrera2018} C.R. Cabrera, L. Tanzi, J. Sanz, B. Naylor, P. Thomas, P. Cheiney, and L. Tarruell, Science \textbf{359}, 301 (2018).
\bibitem{Semeghini2018} G. Semeghini, G. Ferioli, L. Masi, C. Mazzinghi, L. Wolswijk, F. Minardi, M. Modugno, G. Modugno, M. Inguscio, and M. Fattori, Phys. Rev. Lett. \textbf{120}, 235301 (2018).
\bibitem{note1} A more detailed study of the trapping effects will be given in a subsequent paper, with a phase diagram of the mixtures in terms of the chemical potentials \cite{Schaeybroeck2013}.
\bibitem{Lopes2017} R. Lopes, C. Eigen, N. Navon, D. Cl{\'{e}}ment, R.P. Smith, and Z. Hadzibabic, Phys. Rev. Lett. \textbf{119}, 190404 (2017).
\bibitem{Mukherjee2017} B. Mukherjee, Z. Yan, P.B. Patel, Z. Hadzibabic, T. Yefsah, J. Struck, and M.W. Zwierlein, Phys. Rev. Lett. \textbf{118}, 123401 (2017).
\bibitem{Hryhorchak} O. Hryhorchak, and V. Pastukhov, arXiv preprint arXiv:1904.12351 (2019).



\end{thebibliography}
\end{document}